\documentclass[10pt, conference, letterpaper]{IEEEtran}

\ifCLASSINFOpdf
\else
\fi

\usepackage[colorlinks, citecolor=blue]{hyperref}
\usepackage{balance}

\usepackage{setspace, amsmath, amssymb, url, lscape, subfig, algorithmic, multirow, pslatex, listings, verbatim, alltt, amsfonts, wrapfig, boxedminipage, color, cite,bookmark}
\usepackage[vlined,linesnumbered,ruled,boxed]{algorithm2e}
\usepackage[svgnames]{xcolor}
\usepackage{framed}
\definecolor{shadecolor}{named}{LightGray}

\usepackage{paralist}
\usepackage[dvips]{graphicx}
\usepackage{epsfig}
\let\labelindent\relax
\usepackage[inline]{enumitem}

\usepackage{xcolor}

\usepackage{listings}

\makeatletter

\usepackage{listings}

\newcommand\YAMLcolonstyle{\color{red}\mdseries}
\newcommand\YAMLkeystyle{\color{black}\bfseries}
\newcommand\YAMLvaluestyle{\color{blue}\mdseries}

\makeatletter

\newcommand\language@yaml{yaml}

\expandafter\expandafter\expandafter\lstdefinelanguage
\expandafter{\language@yaml}
{
  keywords={true,false,null,y,n},
  keywordstyle=\color{darkgray}\bfseries,
  basicstyle=\ttfamily,                                 
  sensitive=false,
  comment=[l]{\#},
  morecomment=[s]{/*}{*/},
  commentstyle=\color{purple}\ttfamily,
  stringstyle=\YAMLvaluestyle\ttfamily,
  moredelim=[l][\color{orange}]{\&},
  moredelim=[l][\color{magenta}]{*},
  moredelim=**[il][\YAMLcolonstyle{:}\YAMLvaluestyle]{:},   
  morestring=[b]',
  morestring=[b]",
  literate =    {---}{{\ProcessThreeDashes}}3
                {>}{{\textcolor{red}\textgreater}}1     
                {|}{{\textcolor{red}\textbar}}1 
                {\ -\ }{{\mdseries\ -\ }}3,
}

\lst@AddToHook{EveryLine}{\ifx\lst@language\language@yaml\YAMLkeystyle\fi}
\makeatother

\newcommand{\qed}{\nobreak \ifvmode \relax \else
      \ifdim\lastskip<1.5em \hskip-\lastskip
     \hskip1.5em plus0em minus0.5em \fi \nobreak
      \vrule height0.75em width0.5em depth0.25em\fi}

\newcommand{\eg}{{e.g., }}

\newcommand{\comments}[1]{}
\newcommand\hl{\bgroup\markoverwith
  {\textcolor{yellow}{\rule[-.5ex]{2pt}{2.5ex}}}\ULon}

\newcommand{\name}{OaaS\xspace}

\usepackage{orcidlink}

\begin{document}

\title{Tutorial: Object as a Service (OaaS) Serverless Cloud Computing Paradigm}

\author{
    Pawissanutt Lertpongrujikorn\,\orcidlink{0009-0003-4106-2347}, Mohsen Amini Salehi\,\orcidlink{0000-0002-7020-3810} 
     \\ High Performance Cloud Computing (\href{https://hpcclab.org}{HPCC}) Lab, University of North Texas
     \\ $pawissanuttlertpongrujikorn@my.unt.edu,\space mohsen.aminisalehi@unt.edu$ 
    \\[-3.0ex]
}



\maketitle
\IEEEpeerreviewmaketitle

\begin{abstract}
While the first generation of cloud computing systems mitigated the job of system administrators, the next generation of cloud computing systems is emerging to mitigate the burden for cloud developers—facilitating the development of cloud-native applications. This paradigm shift is primarily happening by offering higher-level serverless abstractions, such as Function as a Service (FaaS). Although FaaS has successfully abstracted developers from the cloud resource management details, it falls short in abstracting the management of both data (i.e., state) and the non-functional aspects, such as Quality of Service (QoS) requirements. The lack of such abstractions implies developer intervention and is counterproductive to the objective of mitigating the burden of cloud-native application development. To further streamline cloud-native application development,  we present Object-as-a-Service (OaaS)---a serverless paradigm that borrows the object-oriented programming concepts to encapsulate application logic and data in addition to non-functional requirements into a single deployment package, thereby streamlining provider-agnostic cloud-native application development. We realized the OaaS paradigm through the development of an open-source platform called Oparaca. In this tutorial, we will present the concept and design of the OaaS paradigm and its implementation---the Oparaca platform. Then, we give a tutorial on developing and deploying the application on the Oparaca platform and discuss its benefits and its optimal configurations to avoid potential overheads.
\end{abstract}

\section{Introduction}
\label{sec:intro}

The emergence of cloud technology has drastically transformed the application development process. With cloud infrastructure, provisioning can now be done in a few minutes, as opposed to the weeks or months it used to take. Over the past decade, cloud services have replaced mundane tasks with software automation. The current state-of-the-art, serverless platform utilizes the function-as-a-service (FaaS) paradigm to enable developers to build applications by simply writing code in the form of a function and uploading it to the platform. The system then automates the process of building, deploying, and auto-scaling the application, making the overall development process more effortless and mitigating the burden for programmers and cloud solution architects. 
Major public cloud providers offer FaaS services (\eg AWS Lambda, Google Cloud Function, Azure Function), and several open-source platforms for on-premise FaaS deployments are emerging (e.g., OpenFaaS, Knative).
In the backend, the serverless platform hides the complexity of resource management and deploys the function seamlessly in a scalable manner. FaaS is proven to reduce development and operation costs via implementing scale-to-zero and charging the user in a pay-as-you-go manner. Thus, it aligns with modern software development paradigms, such as CI/CD and DevOps \cite{bangera2018devops}.

\begin{figure}[tbp]
  \centering
  \subfloat[Function as a Service (FaaS)]{\includegraphics[width=0.36\textwidth]{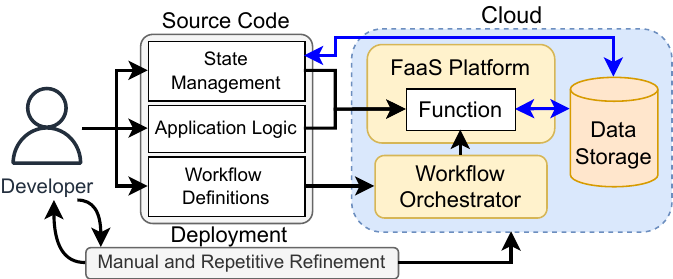}\label{fig:faas_cncpt}}
  \medskip
  \vspace{-6mm}
  
  \subfloat[Object as a Service (OaaS)]{\includegraphics[width=0.42\textwidth]{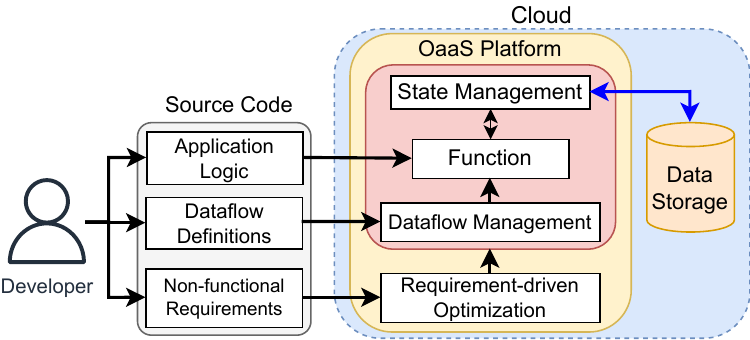}\label{fig:oaas_cncpt}}
  \caption{\small{A bird-eye view of FaaS vs. OaaS. 
  }}
  \label{fig:intro}
  \vspace{-6mm}
\end{figure}    
 
As the FaaS paradigm is primarily centered around the notion of stateless \emph{functions}, it naturally does not deal with the \emph{data}. However, in practice, most use cases need to maintain some form of (structured or unstructured) state and keep them in the external data store. Thus, often the developers have to intervene and undergo the burden of managing the application data using separate cloud services (\eg AWS S3).
For instance, in a video streaming application~\cite{denninnart2024smse}, developers must maintain video files, metadata, and access control in addition to developing functions.  
    

Apart from the lack of data management, current FaaS abstractions do not natively support function workflows. To form a workflow, the developer has to generate an event that triggers another function in each function. However, configuring and managing the chain of events for large workflows becomes cumbersome. Although function orchestrator services 
(\eg AWS Step Function and Azure Durable Function)
can be employed to mitigate this burden, the lack of built-in workflow semantics (see Figure~\ref{fig:intro}) in FaaS forces the developer to intervene and employ other cloud services to chain the functions and navigate the data throughout the workflow manually. In sum, although FaaS makes the resource management details transparent from the developer's perspective, it does not do so for the data, access control, and workflow.

Last but not least, FaaS has limited performance control support. Because the cloud provides separate service abstractions for computing, databases, and other related components (e.g., workflow, messaging, etc.), it prevents the opportunity for the whole application optimization (e.g., data locality, caching, etc.). Moreover, the cloud lacks coordination between the cloud and developers. As a result, cloud service is operated with little knowledge of the application, and developers are less capable of controlling or ``hinting" the system to satisfy the QoS requirements.
\section{Object as a Service (OaaS) Paradigm}
\label{sec:oaas}

To overcome these inherent problems of FaaS, we develop a new paradigm on top of the function abstraction that mitigates the burden of resource, data, and workflow management from the developer's perspective. \emph{We borrow the notion of ``object'' from object-oriented programming (OOP) and develop a new abstraction level within the serverless cloud, called \textbf{Object as a Service (OaaS)}} paradigm~\cite{lertpongrujikorn2023object} to enable cloud-native application developers to unify their logic and data within a single abstraction. 

As shown in Figure \ref{fig:intro}, unlike FaaS, OaaS segregates the state management from the developer's source code and incorporates it into the serverless platform to make it transparent from the developer's perspective. Each application is defined as a collection of cloud objects where its data (a.k.a. state) is modeled as ``attributes'' with supported data types in current cloud data abstraction, and its logic is modeled as methods realized by serverless functions.
In this manner, OaaS abstraction alone is sufficient for the entire cloud-native application development phase---eliminating the need for multiple distinct abstractions and the complexities of effectively gluing them.

\subsection{Optimization Opportunities}
OaaS offers the notions of inheritance and polymorphism to establish software reuse across cloud objects~\cite{denninnart2021harnessing}, thereby avoiding redundancy and enhancing development productivity. Beyond these, OaaS transformation unlocks new opportunities to perform deployment optimizations that would have been difficult, if not impossible, without it. This is because the object abstraction provides richer information for optimization and grants the cloud more control over the deployment to exploit them. For example, in FaaS, stateless function execution is decoupled from its data, making it difficult to minimize the data transmission overhead of functions. In OaaS, however, application data and logic are encapsulated and managed under object abstraction. Thus, OaaS can easily find the data associated with each method and proactively distribute them across the platform instances close to the deployed method, thereby minimizing the data transmission overhead.


\subsection{Dataflow abstraction}

OaaS incorporates dataflow abstraction with built-in data navigation between functions. The significant difference between dataflow abstraction and conventional FaaS workflows is that dataflow introduces the execution flow via the flow of data rather than the invocation order. With dataflow abstraction, the platform handles parallelism and data navigation in the background, reducing developer work and introducing knowledge of data dependency between function invocations for further optimization. Also, developers can change the flow of invocation without changing the function code, only by changing the dataflow definitions.

\subsection{Non-functional requirements interface}

Within the OaaS abstraction, a non-functional requirement interface can be included that lets the developer express their non-functional requirements in a human-friendly manner. Through the interface, developers can declare their non-functional requirements for a whole object or even for a specific part (method). The requirements are defined as high-level and measurable metrics either in the form of QoS (e.g., availability and throughput) requirements or deployment constraints (e.g., budget and jurisdiction). During the deployment, the cloud provider takes these non-functional requirements as input to its internal services and adjusts their operations to meet the requirements. The benefits are three-fold: 
\begin{itemize}[leftmargin=*, itemsep=0pt, topsep=2pt]
    \item \textit{Productivity}: Developers no longer need to handle low-level resource configurations for non-functional requirements, improving productivity by simplifying the deployment process.
    \item \textit{Portability}: incorporating the non-functional requirement interface with OaaS unlocks portability for cloud-native applications. That is, as long as the cloud provider supports OaaS, the application can rely on the object abstraction to maintain its functionality, meet its QoS and constraint expectations (via the non-functional requirement interface), and comfortably migrate across different cloud environments. 
    \item \textit{Cloud-application symbiosis}: The interface fosters a cooperative relationship between the cloud and application developers. It provides cloud providers with clear optimization guidelines to prevent negative impacts on applications and offers developers a way to configure performance and quality without extensive trial and error.
\end{itemize}


\subsection{Potential usages of OaaS}

Typically, applications with unpredictable on-demand workloads are most suitable for a serverless platform since they can fully benefit from an auto-scaling system~\cite{eismann2020review}. This kind of application is also ideal for OaaS if it has the application data to manage. For example, a multimedia processing application that gets triggered when customers upload their files to cloud storage. With FaaS, developers need to work with at least two cloud services (FaaS and cloud storage), but developers only need a single cloud service with OaaS. Suppose developers want to scale their service to broader audiences, they may work on configuring the cloud services in multiple regions, which can be challenging to manage and optimize services since both services are separated. With OaaS, developers explicitly define the relation between data and computation to the platform, which can be used to guide the optimization. The portability aspect of OaaS can also be exploited to streamline application deployment on multiple regions.

Our current Object as a Service (OaaS) design aims to abstract data and functional and non-functional requirements into an object. However, the concept of object abstraction can be extended to provide even greater benefits. For example, we can treat the IoT device as an object that exposes various functions for reconfiguring or accessing the device's capabilities. Consolidating IoT management within a single platform simplifies integration with other parts of the application and streamlines management operations, ultimately enhancing the overall efficiency and versatility of the system.
\section{Oparaca: OaaS-Based Serverless Platform}
\label{sec:oparaca}

To offer the \name paradigm, we develop \textbf{Oparaca} (\underline{\textbf{O}}bject \underline{\textbf{Para}}digm on Serverless \underline{\textbf{C}}loud \underline{\textbf{A}}bstraction) platform. In this section, we will discuss the noteworthy key features of Oparaca.

\subsection{Streamlining the application development}
\begin{figure} [t]
    \centering
    \includegraphics[width=\linewidth]{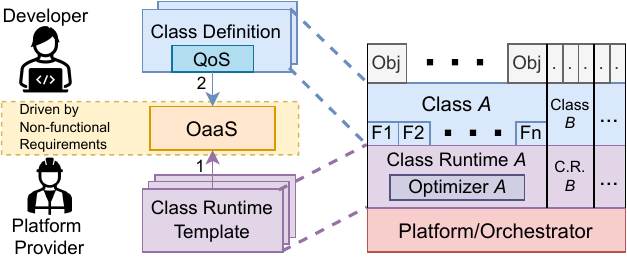}
    \caption{Realizing objects with class runtime and template: OaaS maintains templates customized for various deployment scenarios. For a specific class, Oparaca uses one of its predefined templates to create a class runtime to manage the deployed classes optimally.}
    \label{fig:high-level-oparaca}
    \vspace{-6mm}
\end{figure}

First, Oparaca offers the \emph{class-based} development interface to define the entities of their cloud-native application and non-functional requirements akin to OOP concepts. To that end, the cloud-native application is built on the foundation of \textit{classes}. Each class defines the structure of independent executable objects that are responsible for carrying out one or multiple functionalities. Upon deployment, Oparaca allocates appropriate cloud resources to realize the corresponding objects of the class by creating the \textit{class runtime} (Figure~\ref{fig:high-level-oparaca}) to handle workloads. Moreover, Oparaca supports \textit{inheritance} and \textit{polymorphism} for its classes.

\subsection{Requirement-driven optimization}

Oparaca provides developers with a non-functional requirement interface to guide the performance optimization of their applications in high-level abstraction. This is achieved by allowing developers to define their non-functional requirements. To meet the requirements, Oparaca connects the runtime to the monitoring system and reacts to changes in workload or performance by adjusting the allocated resources or system configuration.

To fulfill the variety of Non-functional requirements on different applications or classes, having the \textit{class runtime} shared among them is difficult to manage because of possible requirements conflicts. To resolve the problem, Oparaca introduces \textit{class runtime template}, 
which provides a configurable class runtime design optimized for a specific set of requirement combinations. When deploying a class, Oparaca will choose from the list the most suitable template to realize the class requirement and then follow the template design to create a dedicated class runtime for this class. Using this approach, Oparaca can make the class runtime have specific characteristics based on the requirement. 
Oparaca also allows platform provider to customize the template configurations, selection conditions, and priority for their operation objective (e.g., resource utilization).

\subsection{Modular and platform-agnostic designs}
Oparaca is designed to be modular and platform-agnostic. Oparaca doesn't tightly rely on any FaaS system or underlying platform but instead uses the standardized API/protocol as the abstraction layer. The most important aspect is that it abstracts developers' code from the cloud storage. Class runtime of Oparaca utilizes the semantic of \emph{pure function} that bundles the object state and input request into the standalone invocation task for offloading this task to the code execution runtime (FaaS engine) and expects the runtime to return with the modified state. Therefore, the code execution runtime is entirely decoupled from the state management. By using an RPC request for offloading a task, any FaaS engine can accept this task to process and return the output and modified state in the response body. Although Oparaca currently only provides comprehensive integration with Knative~\cite{knative}, connecting the other FaaS engine can be done by configuring the URL.


\subsection{Unstructured data support}

Other than the \textit{structured data} (e.g., JSON) supported by the previously-mentioned \emph{pure function} schematic, Oparaca allows developers to combine the \textit{unstructured data} (e.g., multimedia file) as a part of an object state. To meet the platform-agnostic objective, Oparaca uses the S3 protocol~\cite{aws_s3},
a standardizing protocol for object storage for implementing the data access. This approach is not limited to AWS and can be implemented using open-source solutions like MinIO
and Ceph,
which support S3 API. Oparaca employs the \textit{presigned URL technique} to directly allow the developer's code access to the file in object storage without sharing the secret key and avoiding leaking sensitive information.

\section{OaaS Tutorial}

This tutorial aims to instruct the notion of serverless objects and the OaaS paradigm. In addition, we will show how to install and use Oparaca to develop and deploy a cloud-native application. We design the tutorial to consist of the following steps:

\hspace{2mm}
\begin{minipage}{0.95\linewidth}
 \linespread{0.78}
\begin{lstlisting}[
    language=yaml,
    caption=A simplified version of the YAML class definition for image processing,
    label=lst:image-yaml
]
classes:
  - name: Image
    qos:
        throughput: 100  #rps
    constraint:
        persistent: true
    keySpecs:
      - name: image #File Image;
    functions: 
      - name: resize
        #container image
        image: img/resize
      - name: changeFormat
        image: img/change-format
  - name: LabelledImage
    parent: Image
    functions:
      - name: detectObject
        image: img/detect-object
\end{lstlisting}
\end{minipage}

\begin{enumerate} [leftmargin=*]
    \item \textbf{Installing the Oparaca platform.} In this tutorial, we use the local Kubernetes~\cite{k8s} as the container orchestrator and then install Oparaca on top of it. 
    
    \item \textbf{Accessing and managing Oparaca.} Oparaca includes the CLI to facilitate the Oparaca API interaction. This CLI can be used to manage the deployment, access the deployed object, and invoke the function on the object.
    
    \item \textbf{Creating a new function.} Oparaca is designed to work with any framework that accepts and replies to HTTP requests. In this tutorial, we use Python
    code.
    \item \textbf{Defining a new class definition.} Oparaca allows the developer to define the class in JSON or YAML. Listing~\ref{lst:image-yaml} shows The example of defining \texttt{Image} and \texttt{LabelledImage} class. In the class definition, developers have to define the state (lines 7-8) and functions (lines 10-14, 18-19). The non-functional requirements (lines 3-6) are optional.
    
    \item \textbf{Deploying class and interacting with objects.} After creating the class definition, developers can use the CLI command to deploy it to the Oparaca platform. Oparaca then processes the definition to deploy the class runtime. Developers can use CLI, REST API, or gRPC ~\cite{grpc}
    to interact with objects.
    
    \item \textbf{Optimizing the deployment (\textit{class runtime}).} Developers can guide the optimization by configuring the non-functional requirements in the class definition.
\end{enumerate}

The source code, documents, example applications, and deployment scripts of Oparaca are publicly available at the GitHub repository: \url{https://github.com/hpcclab/OaaS}  

\section{Evaluation}

\begin{figure} [ht]
    \centering
    \includegraphics[width=0.75\linewidth]{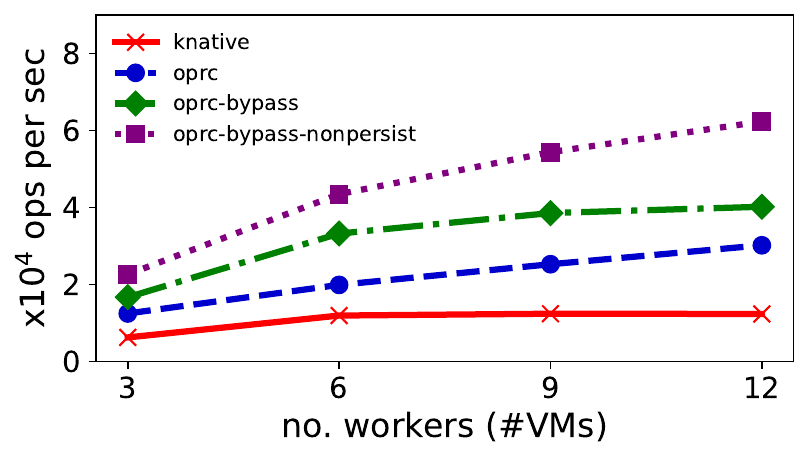}
    \caption{Evaluating the scalability of Oparaca against Knative baseline in JSON randomization application. }
    \label{fig:evlt:scalability}
    \vspace{-6mm}
\end{figure}

In one of our experimental evaluations, we study the scalability of Oparaca by scaling out the workers from 3---12 VMs. 
We compare Oparaca (\textit{oprc}) with Knative as a baseline and add two other versions of Oparaca: \textit{oprc-bypass} that uses a standard Kubernetes deployment as its underlying function execution instead of Knative; Second is \textit{oprc-bypass-nonpersist} that only keeps object data in memory to measure if Oparaca is not bottlenecked by the database write operation. According to Figure~\ref{fig:evlt:scalability}, the throughput of Knative plateaus after reaching 6 VMs is attributed to the database write operation throughput bottleneck. Conversely, Oparaca exhibits the potential for higher throughput due to its reliance on the distributed in-memory hash table to consolidate data for batch write operations. 
Despite not showcasing linear scalability due to the database write performance limitations, Oparaca significantly improves maximum throughput compared to traditional FaaS systems.
\section{Conclusion and Future Work}
\label{sec:conclusion}

The Object-as-a-Service (OaaS) paradigm introduces a new cloud service abstraction that applies principles of object-oriented programming to combine application logic, data, and non-functional requirements into a single deployment package. This approach simplifies native-cloud application development and facilitates requirements-driven coordination among cloud developers, creating opportunities for performance optimization. 
In the tutorial, we demonstrate how to install Oparaca, prototype of OaaS, and develop applications with it.
In the future,  We plan to develop Oparaca to support application deployment across multiple data centers, thereby unlocking the opportunity for non-functional requirements such as latency and jurisdiction.

%
\bibliographystyle{plain} 
\balance
 \bibliography{references}


\end{document}